\documentstyle[12pt,aaspp4]{article}

\newcommand{\ts}{\thinspace}
\newcommand{\p}{$^{\prime}$\ts}

\newcommand{\degree}{\arcdeg}
\def \msun         {\hbox{M$_\odot$}\ }
\def \msunns       {\hbox{M$_\odot$}}
\def \micron       {$\mu$m\ }

\def \lsun         {\hbox{{\it L}$_\odot$}\ }
\def \lsunns       {\hbox{{\it L}$_\odot$}}

\lefthead{Surace et al.}
\righthead{High-Resolution Imaging of Cool ULIGS}

\begin{document}

\title{High Resolution Optical/Near-Infrared Imaging of Cool Ultraluminous Infrared 
Galaxies}

\author{Jason A. Surace}
\affil{Infrared Processing and Analysis Center, MS 100--22, California Institute of 
Technology, Jet Propulsion Laboratory, Pasadena, CA 91125 \\
and \\
University of Hawaii, Institute for Astronomy, 2680 Woodlawn Dr., 
Honolulu, HI, 96822 \\
Electronic mail: jason@ipac.caltech.edu \\} 
\author{D. B. Sanders}
\affil{University of Hawaii, Institute for Astronomy, 2680 Woodlawn Dr., 
Honolulu, HI, 96822 \\
Electronic mail: sanders@ifa.hawaii.edu \\
and\\}
\author{A.S. Evans}
\affil{Dept. of Physics and Astronomy, SUNY, Stony Brook, NY 11794-3800\\
and \\
California Institute of Technology 105-24, Pasadena, CA 91125\\
Electronic mail: aevans@vulcan.ess.sunysb.edu}

\ 

\centerline{Accepted for Publication in {\it The Astrophysical Journal}}

\begin{abstract}

We present high spatial resolution (FWHM $\approx$ 0.3--0.8\arcsec ) BIHK\p-band 
imaging of a sample of ultraluminous infrared galaxies (L$_{\rm ir} >10^{12}$ 
\lsunns; ULIGs) with ``cool'' mid-infrared colors ({\it f}$_{\rm 25\mu m}$/{\it 
f}$_{\rm 60\mu m} < 0.2$) which select against AGN-like  systems and which form a 
complementary sample to the ``warm'' ULIGs of Surace et al. (1998). We find that all of 
the cool ULIGs are either advanced mergers or are pre-mergers with evidence for still-
separate nuclei with separations greater than 600 pc. Extended tidal features such as 
tails and loops as well as clustered star formation are observed in most systems. This 
extended tidal structure suggests a common progenitor geometry for most of the ULIGs: a 
plunging disk collision where the disks are highly inclined with respect to each other. 
The underlying host galaxies have H-band luminosities of 1--2.5 {\it L}$^*$, very 
similar to that found in the ``warm'' ULIGs. The nuclear regions of these galaxies have 
morphologies and colors characteristic of a recent burst of star formation mixed with 
hot dust and mildly extinguished by {\it A}$_{\rm V}$=2--5 magnitudes; only in one 
case (IRAS{\ts}22491$-$1808) is there evidence for a compact emission region with 
colors similar to an extinguished QSO. Most of the observed star-forming knots appear to 
have very young (10 Myr) ages based on their optical/near-infrared colors. These 
star-forming knots are insufficiently luminous to typically provide more than 10\% of 
the high bolometric luminosity of the systems.
\end{abstract}


\section{Introduction}

One of the most important results from the {\it Infrared Astronomical 
Satellite}\footnote{The Infrared Astronomical Satellite was developed and 
operated 
by the US National Aeronautics and Space Administration (NASA), the 
Netherlands 
Agency for Aerospace Programs (NIVR), and the UK Science and Engineering 
Research 
Council (SERC).} ({\it IRAS}) all-sky survey was the discovery of a 
significant population of galaxies that emit the bulk of their luminosity in the 
far-infrared (e.g. Soifer et al. 1984). Studies 
of the properties of these ``infrared galaxies'' showed systematic trends 
coupled to the total 
far-infrared luminosity; more luminous systems were more likely to appear to 
be 
merger remnants or interacting pairs, and were more likely to possess AGN-like 
emission line features.  A more complete review of the properties of 
luminous infrared galaxies is given by Sanders \& Mirabel (1996). 
Much attention has been focused on so-called ultraluminous infrared galaxies 
(ULIGs), 
objects with infrared luminosities, $L_{\rm ir}$,\footnote{$L_{\rm ir} \equiv$ 
L(8--1000\micron) 
 is computed using the flux in all four {\it IRAS} bands according to the 
prescription 
given by Perault (1987); see also Sanders \& Mirabel (1996).  Throughout this 
paper we 
use $H_{\rm o}$ = 75 km s$^{-1}$Mpc$^{-1}$, $q_{\rm o}$ = 0.5 (unless 
otherwise noted).} greater than 
$10^{12}\ L_{\sun}$, which corresponds to the bolometric luminosity 
of QSOs 
\footnote{Based on the bolometric conversion {\it L}$_{\rm bol}$ = 16.5 $\times 
\nu${\it L}$_{\nu}$(B) of Sanders et al.(1989) for PG QSOs. Elvis et al. (1994) 
indicates a value of 11.8 for UVSX QSOs, increasing {\it M}$_{\rm B}$ to -22.5.}
(assuming the blue luminosity criterion $M_{\rm B} < -22.1$, adjusting for our
adopted cosmology: Schmidt \& Green 1983). Multiwavelength 
observations of a complete sample of 10 ULIGs led Sanders et al. (1988a) to 
suggest that these objects might plausibly represent the initial dust-
enshrouded 
stage in the evolution of optically selected QSOs, and that the majority, if 
not all, QSOs may begin their lives in such an intense infrared phase.

Considerable attention has been devoted to so-called ``warm'' systems, which have mid-
infrared colors
$f_{25}/f_{60} > 0.2$.
\footnote{The quantities $f_{12}$, 
$f_{25}$, $f_{60}$, and $f_{100}$ represent the {\it IRAS} flux densities in Jy 
at 12{\ts}\micron, 25{\ts}\micron, 60{\ts}\micron, and 100{\ts}\micron\ 
respectively.} Sanders et al. (1988b) found that these systems predominantly have AGN 
optical spectra, very large molecular gas masses, (M$_{\rm H_2} \sim 10^{10} 
M_\odot$), and advanced merger morphologies, and postulated that they represented the 
immediate transition phase between ULIGs and optically-selected QSOs.

An examination of deep far-infrared flux-limited samples such as the $f_{60} = 1$~Jy 
survey (Kim \& Sanders 1998), however, shows that the majority (80\% ; 90/115) 
of ULIGs are ``cool'' systems (i.e., {\it f}$_{25\mu m}$/{\it f}$_{60\mu m} < 0.2$). 
Thus, these galaxies are similar to the majority of ULIG systems previously studied by 
others (i.e. Sanders et al. 1988a, Kim 1995), rather than the smaller fraction of 
``warm'' AGN-like systems like those discussed by Sanders et al. (1988b), Surace et al. 
(1998; hereafter Paper I) and Surace \& Sanders (1999a; hereafter Paper II). Results 
derived in this paper for a sample of ``cool'' ULIGs are therefore likely to reflect the 
properties of ULIGs as a whole. In Paper II it was discussed that the warm ULIGs were 
possibly a transition state between cool ULIGs and QSOs. If this is true, then the cool 
ULIGs are expected to have properties similar to, yet less evolved than, the warm 
sample.

Many of the cool ULIGs have been imaged before at optical (Sanders et al. 1988, Kim 
1995, Murphy et al. 1996) and near-infrared (Carico et al. 1990, Kim 1995, Murphy 
et al. 1996) wavelengths. However, these observations suffered from poor spatial 
resolution (FWHM $\geq$ 1.0\arcsec) and lack of depth. Their wavelength coverage was 
limited predominantly to R and K-band, and was insufficient to disentangle reddening 
effects from intrinsic colors. Finally, several of the objects in the cool sample presented 
here have never been imaged before. 

We present here new multiwavelength observations with 1.5 and 4$\times$ the spatial 
resolution of previous ground-based observations at optical and near-infrared 
wavelengths; despite being ground-based, they allow us to isolate interesting features 
such as the star-forming knots detected in the warm ULIG sample. 

\section{The Sample}

A sample of 18 ``cool"
(i.e., {\it f}$_{\rm 25\mu m}$/{\it f}$_{\rm 60\mu m} < 0.2$)
ULIGs was drawn from the {\it IRAS} Bright Galaxy Sample of
Sanders et al. (1988) as well as the IRAS 1-Jy sample (Kim \& Sanders
1998).  The cool ULIG sample was chosen to complement the samples of
``warm" ULIGs and infrared-excess PG QSOs as part of a study of the
possible evolutionary connection between ULIGs and optically-selected
QSOs (Surace 1998).  A key observational fact known from spectroscopic
studies of the larger parent samples is that cool ULIGs, warm ULIGs,
and PG QSOs are known to represent a spectroscopic sequence that ranges
from objects whose distribution of spectral types is biased towards
H~II-like spectra (HII-40\%, LINER-40\%, Sy2-20\%; e.g. Veilleux, Kim,
\& Sanders 1999), to objects dominated by Seyferts (HII-10\%, LINER-20\%,
Sy2-40\%, sy1-30\%; Veilleux et al. 1997), and finally to optically-selected
Sy~1s (Schmidt \& Green 1983; i.e. part of the definition of QSOs).
The 18 cool ULIGs discussed here have a similar distribution of
spectral types (HII-54\%, LINER-35\%, Sy2-11\%)as their parent sample.
The ``warm'' sample has been discussed previously in Papers I \& II,
and the infrared-excess PG QSOs are the subject of a forthcoming paper.

All of the ``cool'' ULIGs have been chosen to lie within the volume
{\it z} $<$ 0.16. This is the same volume limit as the ``warm'' ULIG
sample of Paper I, and is very close to the completeness limit for ULIGs
in deep {\it IRAS} surveys. Also, this is sufficiently nearby that the
spatial resolution achievable from the ground can provide information on
scales known from Papers I \& II to be physically meaningful (typically
a few hundred parsecs). Since there are over 100 such ULIGs known,
this sample was selected first to include the original 7 cool ULIGs
in the BGS sample, as these are most well-studied.  The remaining cool
ULIGs were selected such that their redshift distribution was similar
to the ``warm'' ULIG and PG QSO samples of Sanders et al. (1988b) and
Surace (1998).  Specific objects were randomly chosen to lie in regions
of the sky more amenable to observation from Mauna Kea, and to ameliorate
crowding of the observing program in spring. Since this selection criterion
is unrelated to the physical properties of the ULIGs, it should not bias
the sample.

\section{Observations and Data Reduction}

The data were taken between October 1995 and March 1998 at the f/31 focus of the UH 
2.2m telescope using a fast tip/tilt image stabilizer. This image stabilizer consists of a 
piezo-driven secondary with a pick-off mirror and guider ccd  for guide-star 
acquisition. It is described by Jim (1995) and Pickles et al. (1994), and was used for 
the observations of Papers II \& III. When used in off-axis guiding mode, it eliminates 
common-mode vibration of the telescope, as well as some seeing effects. This results in 
near-diffraction limited imaging (FWHM $\approx$ 0.3\arcsec ) in the near-infrared 
most of the time, since the seeing at the UH 2.2m site is extremely good. The system is 
not as effective at optical wavelengths, where the seeing is much poorer. The spatial 
resolution at I-band is typically 0.5-1\arcsec, with 0.75\arcsec \ being typical. At B-
band it is usually 1\arcsec .

The near-infrared data were taken with the QUIRC 1024$\times$1024 HgCdTe camera 
in a manner identical to that of the ``warm'' ULIG sample in Paper II. 
The observations were made at H (1.6$\mu$m) and K\p (2.1$\mu$m).
The H filter was chosen since it is the longest wavelength filter which is still
 relatively unaffected by thermal dust emission; dust hot enough to emit significantly at 
this
wavelength would be above the dust sublimation temperature. The choice of the 
University of Hawaii K\p filter, which is bluer than both the Johnson K and the 2MASS 
K$_S$, was motivated by the lower thermal sky background in the K\p-band (Wainscoat 
\& Cowie 1992). This improves the detectability of faint features such as star-forming 
knots. Throughout this paper we exclusively refer to K\p. Comparison to work by other 
authors is made using the conversion of Wainscoat \& Cowie (1992). 

The near-infrared data were reduced in the same manner as that described in Paper II. 
The data was initially sky-subtracted using consecutive, dithered frames; because the 
QUIRC field of view is so large (60\arcsec ), it was possible to dither the target on-
chip, thereby increasing telescope efficiency by a factor of 2. The images were then 
flattened using median flats constructed from images of the illuminated dome interior. 
Each image was masked by hand to exclude bad pixels and regions contaminated by 
negative emission introduced by the sky subtraction. The images were aligned using the 
IMALIGN task in IRAF, which uses a marginal centroiding routine that calculates a best 
fit solution to a number of (user-supplied) reference stars in the field. Typical 
alignment errors were estimated (on the basis of the fit) to be about 0.25 pixels. Given 
that the data was typically sampled by 5 pixels FWHM for a point source, alignment 
errors are unlikely to be important. Images were scaled according to their exposure 
times and then, in order to account for any variable sky background, an offset was 
subtracted from each image based on the background measured in that frame. The images 
were combined by medianing using IMCOMBINE and rejecting pixels outside the linear 
regime of the array.

The optical data were taken with several different instrument configurations. The UH 
Tektronix 2048 and Orbit 2048 cameras were used at the f/31 focus of the UH 2.2m. In 
both cases, the data were binned 2$\times$2 in order to provide better spatial sampling 
(with adopted binned pixel sizes of 0.14\arcsec \ and 0.09\arcsec , respectively). The 
Orbit 2048 was also used in conjunction with the HARIS spectrograph; by withdrawing 
the dispersion components it was possible to image through the spectrograph. The 
telescope f/31 beam was reimaged at f/10 resulting in an unbinned image scale of 
0.14\arcsec \ pixel$^{-1}$.

The optical data reduction involved several steps. First, the CCD bias pattern was 
removed by subtracting from each image a high S/N median bias frame constructed from 
sequences of 20-30 bias frames taken at the beginning and end of each night. Pixel-to-
pixel response variations were then corrected by dividing each image by a high S/N flat 
produced by making dithered observations of the twilight sky in each filter. Typical 
twilight exposures were 2-3 seconds each, short enough to avoid getting detectable flux 
from field stars, yet long enough to avoid flat-field errors introduced by the radial 
shutter used at the UH 2.2m. Estimated S/N for the flats (based on Poisson statistics and 
the gain of the CCD) was between 250-500. Neither CCD showed any evidence of 
measurable dark current, based on an examination of long closed shutter exposures. The 
images were then corrected to normal orientation by transposition and rotation using the 
ROTATE task in IRAF based on the known field rotation of the cassegrain focus of the UH 
2.2m, which is accurate to better than 1 degree. The CCD overscan regions were 
trimmed using IMCOPY. The images were shifted and aligned using the method detailed 
above for the near-infrared data. The images were then averaged using an algorithm that 
rejects pixels inconsistent with the known noise properties of the CCD, allowing for 
rejection of cosmic rays. The shifted images were combined onto a larger image than the 
original data frames, thereby increasing the total field of view due to the dithering 
process. This was valuable primarily in order to increase the availability of PSF stars, 
since the size of the camera focal plane was much larger than the measurable extent of 
any of the galaxies.

In some cases the telescope cassegrain focus was rotated during the night in order to 
acquire brighter guide stars, thereby allowing faster guide rates and improving quality 
of the tip/tilt guiding. Additional flats were made, whenever feasible, at the rotation 
angles used. This was necessary since changes in the illumination of dust and other 
defects in the telescope optics tended to produce notable changes in the flat-field 
response, resulting in a strong background gradient in the vicinity of any optical defects 
such as dust.

The data were calibrated through observations of standard stars in the optical and near-
infrared ( Landolt 1983, 1992, Elias et al. 1982). Typically, 4--5 standard stars 
were observed throughout the night at airmasses similar to those of the science targets. 
The stars were interleaved with the science targets; such a strategy also enabled them to 
be used for refocussing the telescope. In most cases the nights were photometric with 
1$\sigma$ uncertainties in the photometric calibration of 0.05 magnitudes. For non-
photometric nights, the data was calibrated by forcing agreement with large-aperture 
photometry already in the literature. In particular, the optical data for UGC 5101 were 
calibrated using Sanders et al. (1988a), while the IRAS 00091-0738 and IRAS 
01199-2307 data were calibrated using the photometry of Kim (1995). K-corrections 
have not been applied to any magnitude reported in this paper. Since the most distant 
object is at redshift {\it z}=0.16, K-corrections are likely to be quite small. For a 10 
Myr old starburst of the sort discussed below at the median redshift, the K-corrections 
are $\delta_M$(B,I,H,K\p)=(0.12,0.13,0.07,-0.11), and for a 2 Gyr old starburst 
population they are $\delta_M$(B,I,H,K\p)=(0.61,0.19,0.08,-0.16). Kim (1995) 
and Trentham et al. (1999) computed K-corrections based on very large aperture 
spectra for ULIGs at redshifts similar to this sample, and found that optical K-
corrections were typically less than 0.25 magnitudes, and near-IR K-corrections were 
less than 0.1 magnitudes. In order to make it easier to compare the measurements in this 
paper to those of others, magnitudes are presented without the uncertainty introduced by 
an assumed spectral shape. While the bandpass compression term is known for each 
ULIG, the inclusion of a bandpass compression-only, ``pseudo K-correction'' is omitted 
since the possible confusion it would create outweighs the very small advantage of 
including it.

The point-spread-function (PSF) was calibrated with actual stars in the final combined 
science image using DAOPHOT in the manner described in Paper II. All of the stars were 
identified, scaled, shifted, and combined using a sigma-clipping algorithm and weighting 
according to total flux, thus creating as high a S/N PSF image as possible. In those few 
cases where no stars were found in the science images, the PSF was estimated by using 
the closest temporally adjacent standard star. Since the tip/tilt guiding has little effect 
on atmospheric distortions at short wavelengths, this technique works well for optical 
data. Similarly, since the seeing remains stable on timescales of many minutes, this 
technique is also effective in the near-infrared.

Additionally, some of the cool ULIGs have been observed  by {\it HST}/WFPC2 through 
the F814W ({\it I}-band) filter as part of the Borne et al. (1997) ULIG snapshot 
survey and are publicly available from the {\it HST} archive (see Table 1). These data 
were reduced in the same manner as the WFPC2 data in Paper I. These data were included 
primarily for comparison with the ground-based data. Since they are only at a single 
wavelength corresponding directly to one of our ground-based filters, they cannot be 
easily used for the multiwavelength color analysis presented here. Their spatial 
resolution is too high to allow a direct comparison without significant aperture effects.

In two cases we do not have complete data. Near-infrared data was not taken of Arp 220 
because it had already been observed with {\it HST/NICMOS} (Scoville et al. 1998); this 
data has been retrieved for use here. Mrk 273 was not observed in the near-infrared 
because it had already been observed with adaptive optics from CFHT (Knapen et al. 
1997).

As in Paper II, in some cases high spatial resolution techniques such as deconvolution 
were applied to the data in order to enhance detectability of features. The Richardson-
Lucy algorithm implemented in IRAF/STSDAS was used along with the data-derived PSFs. 
It was allowed to iterate 20--50 times until noticeable artifacting appeared. Previous 
use of this technique has shown the observed structure to be reliable; in the case of the 6 
systems with data from WFPC2 this was checked directly. Magnitudes and colors were 
not derived from the deconvolved data; rather, the deconvolved data was used to clarify 
morphological details. The actual photometry was measured from the raw data using 
aperture photometry and PSF-derived aperture corrections. For further details, see 
Surace (1998). 

\section{Results}

\subsection{Morphology}

\subsubsection{Large-Scale Features}

Images of each ULIG in the ``cool'' sample in each of the four observed filters are 
presented in Figure 1. Figure 2 presents near-truecolor images constructed from the 
optical data, as by Surace et al. (1998). The B \& I-band images were linearly 
interpolated to provide the color information, and the resulting truecolor images should 
appear similar to the actual colors perceived by the eye. \footnote{Human vision is 
adapted to understand three-dimensional structure in terms of simultaneous luminosity 
and color information, and thus such true-color images provide the most intuitive means 
of understanding the structure of the star-forming regions and the reddening resulting 
from embedded dust (Kandel et al. 1991, Malin 1993).} Despite having higher spatial 
resolution, the near-infrared morphology is more featureless; therefore a near-
infrared color image was not included.

The cool ULIGs exhibit a wide variety of morphologies, the details of which are tabulated 
in Table 2. At least 6 of the 14 systems (43\%) have obvious double galaxy nuclei as 
evidenced by the manner in which tidal tails and spiral structure are centered on high 
surface brightness, extended emission regions. This is similar to the value of 47\% 
found by Murphy et al. (1996) for a sample of 53 non far-infrared color-selected 
ULIGs. However, an additional 4 systems {\it may} have double nuclei, which would 
bring the total double nucleus fraction to as high as 72\% (10/14). 

Projected separations of the definite double nuclei span an order of magnitude from 25 
kpc in IRAS{\ts}01199$-$2307 to just 2.5 kpc in IRAS{\ts}22491$-$1808. In all 
cases the double nucleus systems have extended tidal structure indicating that these 
systems have already passed initial perigalacticon and are now in an advanced merger 
state. None of the systems which appeared to have single nuclei from previous 
observations were found to have double nuclei when observed at higher spatial 
resolution. This may be an effect of confusion. As noted in Paper I and illustrated by the 
cautionary tale of Mrk 231 (Armus et al. 1994), it is difficult to differentiate true 
galactic nuclei (spheroidal bulge remnants) from unresolved aggregates of luminous, 
dusty young star-forming regions. Conversely, at the median redshift of the sample, 
typical spatial resolutions of 0.4\arcsec \ at K\p are $\approx$ 600 pc, so at least a 
few systems with projected separations of 0.5-2 kpc should be expected. Even in the 
most confused of cool ULIGs (IRAS{\ts}22491$-$1808), the two galaxy nuclei can be 
clearly differentiated from the star-forming clusters on the basis of their K\p emission. 
Furthermore, the extremely high spatial resolution  HST observations of three of the 
single nucleus objects also fail to detect additional nuclei. There are several apparently 
single nucleus systems which have extended optical nuclear regions with apparent 
bifurcating dust lanes which are very similar in appearance to the bifurcated core of 
IRAS{\ts}05189$-$2524. It is possible that the original progenitor galaxy nuclei lie 
in this chaotic central region if they have not already coalesced since they are not 
anywhere else in the system, and it seems unlikely that either heavy dust obscuration or 
a remarkable alignment along the line-of-sight could hide them. In the case of the 
nearest such system, Arp 220, the nuclei are actually known to be hidden in the optical 
structure.

The failure to discover any additional systems with previously unknown double nuclei 
when observed with the much higher spatial resolution of the near-infrared 
observations is similar to the result found in Paper I for warm ULIGs newly observed by 
{\it HST}. In order to evolve from the very widely separated systems to the single 
nucleus merger systems, the ULIGs must pass through a stage where the nuclei are 
separated by 0.5-2.5 kpc. If we assume an even distribution of physical separations 
between 25 kpc and fully merged systems, then the probability of selecting at random a 
sample with as few small-separation systems as observed here is much less than 1\% 
(based on monte carlo simulations). This implies a bimodal distribution in separations 
for the underlying distribution of ULIGs. This may indicate that the timescale for final 
coalescence of the nuclei is comparatively short, and as a result there is a natural 
depletion in intermediate separation systems. Alternatively, the ULIG luminosity 
selection criterion may be selecting two different populations --- systems that are 
ultraluminous at large separations, and systems that are ultraluminous only after 
merger. 

None of the cool ULIGs shows evidence for the same kind of bright, extremely compact 
AGN-like nuclei as are found in several of the warm ULIGs such as Mrk 231. However, 
the {\it HST} images of Mrk 273 and UGC 5101 (Figure 3) reveal that both of these 
systems have single, very compact (radius $\approx$ 100 pc) high surface brightness 
features at {\it I}-band in their nuclear regions. Because there is also considerable 
structure in the nuclei of these systems, our {ground-based} {\it I}-band images with 
limited spatial resolution cannot spatially separate these central nuclei from their 
luminous surroundings. Similarly, both our ground-based images and the {\it HST} data 
show the presence of a nearly unresolved object in IRAS 12112+0305. 
IRAS{\ts}23365+3604 also appears to have very compact nuclear structure, although 
it also has one of the lowest percentages of its total emission in its central 2.5 kpc 
region. These results point to the presence of compact ``nuclei'' in a small percentage 
(28\%) of the cool ULIGs that are morphologically similar to the most extended, low 
surface brightness ``nuclei'' in the warm ULIGs, such as IRAS 12071$-$0444 and IRAS 
15206+3342. Three of these four cool ULIGs are single nucleus systems (Mrk 273, 
UGC 5101, and IRAS{\ts}23365+3604), and all four have considerable numbers of 
star-forming knots visible near their nuclei in the {\it HST} data; all of these are 
suggestive of transition objects between the greater population of cool ULIGs (e.g., IRAS 
14348$-$1447) and the AGN-like compact-nucleus warm ULIGs (e.g., Mrk 231). 
That two of these systems (Mrk 273 and UGC 5101) are the only two cool ULIGs with 
known Seyfert spectra makes this explanation even more compelling (Sanders et al. 
1988a, Khachikian \& Weedman 1974).

All 14 systems show well-developed tidal tails and plumes. Many of these tails are 
curved to form circular or semi-circular ring-like structures, and it is likely that 
many of the linear tails are similar structures seen edge-on. The tails have total 
projected lengths averaging 35 kpc and ranging from 9 kpc to 100 kpc; the ringlike 
structures are generally 15-30 kpc in radius. In almost every case, these tidal features 
appear to be circular rings or ring segments oriented nearly perpendicular to each 
other. In addition to these obvious cases where the disks and tails lie in a plane either 
parallel or orthogonal to the projected plane of the sky, many of the other galaxy 
systems can be explained as projections of this same geometry combined with additional 
rotation. In many cases it appears that the progenitor galaxies may have been rotating in 
opposite directions, judging from the opposing opening angles of the tidal tails, which 
have presumably inherited the angular momentum of their progenitors (e.g., IRAS 
01199$-$2307, IRAS 23233+0946). 
Figure 4 shows an n-body simulation taken from Barnes (1992). The upper left panel 
shows the model disk encounter geometry, which may be similar to that of many of the 
cool ULIGs, while the bottom right panel shows the cool ULIG UGC 5101 at I-band for 
comparison.

\subsubsection{Star-Forming Knots and Small-Scale Structure}

Many of the systems show evidence of the same clustered star formation seen in the 
``warm'' ULIGs. Only four show no obvious high surface brightness compact knot-like 
features. Since these are the four systems at the highest redshifts ({\it z} $ > $ 0.13), 
this may be an effect of the limited spatial resolution achievable from the ground; 
however, the physical spatial resolution for IRAS{\ts}00091$-$0738 is nearly as 
poor at {\it z} = 0.12, yet it shows considerable evidence for star-formation. Moreover, 
these systems are 4 of the 6 double nucleus systems and 3 of them have the largest 
projected separations, which may indicate a lack of clustered star formation in mergers 
with large nuclear separations. This is similar to other results which indicate that there 
may be a delay in star formation until some time after the first contact between galaxies 
(Joseph et al. 1984, Surace \& Sanders 1999b). The role of star formation as 
characterized by far-infrared activity in very widely separated pairs is discussed 
widely in the literature (Bushouse et al. 1988, Haynes \& Herter 1988, Surace et al. 
1993). Figure 5 illustrates Richardson-Lucy deconvolved data for these star-forming 
regions. Figure 3 presents images of the {\it HST}/WFPC2 data for our sample objects, 
which can then be compared to the ground-based images in Figure 5. 

In order to be recognized as real features in the deconvolved images, the small-scale 
structure must at least be recognizable in the undeconvolved images. This provides a 
means of allowing the real features to be discriminated from the amplified, highly-
correlated noise that produces the ``mottling'' effect in the deconvolved data. As in Paper 
I, the knots are defined as compact emission sources with closed isophotes that are more 
than 3$\sigma$ above the local background in the undeconvolved images. This distinction 
is made in an attempt to discriminate between the ``knots'', which appear to be compact 
bursts of star-formation, and the more extended ``condensations'' which appear to be a 
result of large-scale tidal structure. Examples of condensations can be seen in the 
southern arc of IRAS 12112+0305 and the western tail of UGC 5101. As was noted in 
Paper I, it is likely that all of the ``knots'' are actually unresolved aggregates of star-
forming clusters like those seen in other, more nearby interacting galaxies. Given the 
even poorer spatial resolution of the ground based images (as opposed to those from {\it 
HST}), many of the star-forming knots are likely to be even more confused than in 
Paper I and this limits our ability to recognize the star-forming knots. Because of this, 
the analyses of the luminosity functions, etc., as was carried out in Paper I, cannot be 
repeated here meaningfully. Table 4 gives aperture photometry for the star-forming 
knots that could actually be recognized as such; it probably misses many more. It also 
does not list photometry for features that appeared to be more extended tidal structure, 
although in many cases there appears to be tidal structure in the form of arms and wisps 
even at very small scales. Details of this additional structure are discussed in \S 4.2. 
Positions are given relative to the brightest feature in I-band and correspond to the 
apparent galaxy ``nucleus''. 

Most of the star-forming knots are within a radius of a few kpc of the nuclei, a result 
suspected from previous aperture photometry studies which showed systematic color 
changes at small galactocentric radii (Carico et al. 1990). As in the warm ULIGs, knots 
and condensations are also seen along the tidal features; this is particularly apparent in 
IRAS{\ts}00091$-$0738, UGC 5101, Mrk 273, IRAS 12112+0305, 
IRAS{\ts}14348$-$1447, and IRAS 22491$-$1808.

\subsection{Luminosities}

The ``cool'' ULIGs do not show morphological evidence for the compact putative AGN found 
in the ``warm'' ULIGs. Instead, most of the ``nuclei'' appear to be extended with evidence 
of starburst activity, based on their colors as detailed below. Furthermore, many of the 
nuclei have complex morphological structures similar to star-forming regions. 
Therefore, only the luminosities of the host galaxies and the star-forming knots are 
discussed since a comparison between the nuclei and AGN is unwarranted.

\subsubsection{Host Galaxies}

Luminosities were computed using the formula given in Paper I (equation 1), which 
corrects for distance but does not include K-corrections. As in Paper II, we consider the 
H-band luminosity as the best indicator of the total mass of the old stellar population, 
and hence the combined mass of the merger progenitors. It additionally is much less 
affected by extinction than shorter wavelength observations, and K-corrections at H-
band are likely to be less than 0.1 magnitude based both on a modeled starburst 
population and also on the empirical results of Kim (1995). Unfortunately, this is 
somewhat more complicated in the case of the cool ULIGs than it was for the warm ULIGs 
and will be for quasar host galaxies (e.g.  Surace (1998), Surace \& Sanders (1999a)) 
. In particular, since the warm ULIGs each appeared to contain a compact AGN-like 
nucleus, it was easy to subtract this nuclear component, as well as any emission from 
compact star-forming knots, from the extended host galaxy. In the case of many of the 
cool ULIGs, there is not a clear AGN component. Instead, the ``nuclei'' of the cool ULIGs 
often appear to be diffuse, extended regions with complex structure; a large fraction of 
the luminosity in the nuclear regions is likely to be old starlight, in which case it should 
not be subtracted from the global H-band luminosity.

Several approaches are considered in computing the total H-band luminosity of the host 
galaxies. First, the total flux at H-band, including any star-forming knots, galaxy 
nuclei, etc. is considered. This will set an upper limit on the host galaxy luminosity, 
since it must necessarily include additional young stellar population components.
The integrated photometry was derived by measuring the total flux of the system in an 
aperture large enough to encompass the optical extent of the galaxy at a flux level below 
1 $\sigma$. Errors in total luminosity are typically 0.07 magnitudes. These values are 
given in Table 3.
In this way it is found that the cool ULIGs are similar in luminosity to the warm ULIGs. 
In Paper II the H-band luminosity of an {\it L}$^*$ galaxy was estimated to be between 
{\it M}$_{\rm H}$=$-$23.8--24.1; we adopt here {\it M}$_{\rm H}$=$-$23.9. 
The cool ULIGs are found to have total luminosities ranging from {\it M}$_{\rm H}$=$-
$23.38 (Arp 220) to {\it M}$_{\rm H}$=$-$24.70 (IRAS{\ts}14348$-$1447, 
IRAS{\ts}22206$-$2715), with a mean value of {\it M}$_{\rm H}$=$-$24.23. This 
range is from 0.6--2.1 {\it L}$^*$, with a mean of 1.4 {\it L}$^*$ and with roughly 
half of the ULIGs lying between 1.5--2.5 {\it L}$^*$. However, there are no 5--7 {\it 
L}$^*$ systems as there were with the warm ULIGs.

This result is complicated by the finding that the colors of the nuclear regions of the cool 
ULIGs are consistent with a line-of-sight extinction of {\it A}$_{\rm V}$ = 2--5 
magnitudes (see \S 4.3). As was indicated in Paper II for IRAS{\ts}08572+3915, it is 
therefore likely that the unreddened global values derived above are slight 
underestimates. One visual magnitude of reddening corresponds to {\it A}$_{\rm H}$ = 
0.18 magnitudes (Reike \& Lebofsky 1985). In the most extreme (hypothetical) case 
that all of the luminosity lay within the inner 2.5 kpc (since these reddening values 
were determined inside this region), then dereddening would at most increase the 
luminosity of the galaxies by 1 magnitude at {\it H}. In actuality this effect is much 
more modest, as can be demonstrated by dereddening just the observed luminosity in the 
nuclear regions and adding this to the outer galaxy luminosity; typically this increases 
the galaxy luminosity by 0.1 magnitudes, ranging from 0.08 in IRAS{\ts}23233+0946 
to 0.4 in UGC{\ts}5101. The dereddened luminosities then range from {\it M}$_{\rm 
H}$=$-$23.8 to {\it M}$_{\rm H}$=$-$24.9 or 0.9--2.5 {\it L}$^*$. Excluding the 
anomalous case of IRAS{\ts}01003$-$2238, this is very similar to the range found for 
the warm ULIGs, and is consistent with the merger of two galaxies, a result also 
suggested by the extended morphologies of the ULIGs.

At the other extreme, it is possible to examine the luminosity of the outer galaxy by 
simply excluding all of the flux in the 2.5 kpc diameter nuclear region. On average, 
37\% of the H-band flux lies within a radius of 1.25 kpc of the center of each ULIG, 
varying from 52\% (IRAS{\ts}00091$-$0738) to 17\% (IRAS{\ts}23365+3604). 
Excluding the central regions, the luminosity of the outer regions of these galaxies range 
from {\it M}$_{\rm H}$=$-$23.0 (0.4 {\it L}$^*$) to {\it M}$_{\rm H}$=$-$24.5 
(1.7 L$^*$). Obviously, these are considerable underestimates --- the previous 
dereddened results are probably closer to the truth. Furthermore, the fraction of H-
band light inside the central regions is similar to that found in the bulges of disk galaxies 
(Kent 1985), further indicating that the central region contains a considerable amount 
of old starlight.

\subsubsection{Star-Forming Knots}

The star-forming knots have {\it B}-band luminosities ranging from {\it M}$_{\rm 
B}$=$-$14.4 to {\it M}$_{\rm B}$=$-$18.5, which is similar to or slightly higher 
than those observed for the ``warm'' ULIGs. The higher values are likely to result from 
confusion. The more limited spatial resolution of the ground-based optical observations 
(typically 0.7--0.8\arcsec at {\it B}) yields a physical resolution of 1 kpc at the 
median redshift. Any given knot optically detected from the ground is therefore likely to 
contain at least several of the knots detected by {\it HST}. An examination of Figure 3 
confirms that in some cases (i.e., Mrk 273) this occurs, while in others (IRAS 
22491$-$1808) this effect may be much less. Also, in Paper I it was found that the 
luminosity function of the knots continued to increase until reaching the detection limit 
of {\it M}$_{\rm B}$=$-$12, but confusion and limited spatial resolution prevent our 
reaching such low detection limits. Finally, any differences in the ages of the stellar 
populations of the knots would be expected to change their luminosities by several 
magnitudes. The total integrated luminosity of the star-forming knots (the sum of all 
knot luminosities) at B-band ranges from {\it M}$_{\rm B}$=$-$17.2 in IRAS 
00091$-$0738 to {\it M}$_{\rm B}$=$-$19.5 in IRAS 14348$-$1447, with a 
median value of {\it M}$_{\rm B}$=$-$18.3. This is approximately 6 times more 
luminous than the mean integrated {\it B}-band luminosity of the knots in the warm 
ULIGs. The total fraction of the galaxy {\it B}-band luminosity found in the star-forming 
knots ranges from 6\% to 25\%, with a median of 11\%. By comparison, the warm 
ULIGs vary from less than 1\% (PKS 1345+12) to nearly 40\% (IRAS 
15206+3342), but this much larger spread in the total percentage is due to the 
presence of the putative active nuclei. The highest percentage (and total luminosity) are 
found in the two double nuclei with detectable star-forming knots: IRAS 22491$-
$1808 and IRAS 14348$-$1447. This may be indicative of the increased luminosity of 
the knots with younger ages, if the presence of double nuclei actually implies younger 
age.

Very few of the knots are detected at near-infrared wavelengths. Significant numbers of 
detections occur only in IRAS 12112+0305, IRAS 15250+3609, and IRAS 22491$-
$1808. The K-band luminosities for the detected knots range from {\it M}$_{\rm 
K^{\prime}}$=$-$18.0 to {\it M}$_{\rm K^{\prime}}$=$-$23.0, which again is 
similar to those found in most of the ``warm'' ULIGs. The implications of the non-
detections are discussed below. The total integrated luminosity of the star-forming knots 
detected at K\p ranges from {\it M}$_{\rm K^{\prime}}$=$-$18.0 to {\it M}$_{\rm 
K^{\prime}}$=$-$23.5. Considering the upper limits imposed by the non-detections, 
the typical cool ULIG has an {\it M}$_{\rm K^{\prime}}$ originating in the star-
forming knots of no more than $-$21.2. In those warm ULIGs with knots detectable in 
the near-infrared, this same number varied from $-$20.6 to $-$24.0, a very similar 
range.

\subsection{Colors}

\subsubsection{Models}

A multi-color approach as used by Surace \& Sanders (1999) is adopted here. The three 
colors ({\it B$-$I}),({\it I$-$H}),({\it H$-$K\p}) define a spectral shape. Two 
representations of this color space are presented in Figures 6 \& 7. These 3-color 
diagrams are the set of all possible SEDs that can be defined (within the upper and lower 
bounds of the axes) by the four photometric measurements.  The tracks made by the 
models (described below) represent the subset of all possible SEDs consistent with those 
models. Thus, the points on the model tracks lying closest to the real data represent the 
best possible fit of the data to the model. The two figures are identical and represent 
different rotations of the color space with respect to the line-of-sight-reddening vector. 
Figure 6 is rotated such that the plane of the page is orthogonal to this vector, and hence 
the location of a point in the projection of the 3-color space onto the page is independent 
of line-of-sight extinction. It is therefore possible to fit the data to the models 
independent of extinction. Figure 7 is rotated such that the vector lies in the plane of the 
page and thus allows an immediate estimate of the magnitude of line-of-sight extinction. 
The value of the two rotated projections is that they reduce the 3-color diagram to the 
more familiar 2-color diagram, only with the special property that they have separated 
the effects of line-of-sight extinction.

Figures 6 \& 7 also show the colors of several modeled populations, as well as various 
reddening effects (see the figure caption).
A more thorough explanation of these models is given by Surace \& Sanders (1999).
For comparison, the median colors of the ``warm'' ULIGs, which for reasons presented in 
Paper II are likely to be AGN viewed along a complex, lightly extinguished path, are 
shown with a large circle.

Rather than K-correct the data with an unknown SED, a different approach was used 
wherein the models were corrected instead. Since the models of the emission processes, 
starbursts, 
and QSOs have detailed SEDs, inverse K-corrections can be made to their rest-frame 
colors at the redshift of our targets. This is done by convolving the synthetic spectra 
with the known detector and filter bandpasses. The magnitude zeropoint calibration of the 
filters is derived using the Kurucz model spectrum of Vega (BC95). For brevity, 
Figures 6 and 7 are calibrated to the median ULIG redshift, z=0.1. The K-corrections 
affect the modeled stellar colors by $\delta$({\it B$-$I,I$-$H,H$-
$K\p})=($-$0.01,0.06,0.18) 
for a young (10 Myr) starburst, and by (0.42,0.11,0.24) for an old (2 Gyr) 
population. The effects of K-corrections can therefore be quite large depending on the 
modeled population, and hence the representation of stellar colors made in Figures 6 \& 
7 can be considered to 
have a sizable uncertainty attached. Larger redshift values also shift the dust emission 
curves to a more vertical orientation, as the rest frame filter bandpasses become bluer. 
Actual comparisons to the models made in the text were performed using models 
corrected to the appropriate redshift.

\subsubsection{Data}

The same confusion over the definition of ``nuclei'' that plagues the determination of the 
underlying galaxy luminosity also creates problems for the color analysis. In order to 
help eliminate redshift-dependence, ``nuclear'' is defined to mean the central region of 
the galaxy 2.5 kpc in diameter. At the redshift of the most distant object in our sample 
({\it z}= 0.152), this corresponds to 0.8\arcsec , which under the worst conditions is 
roughly the size of one resolution element at optical wavelengths. The nuclear 
magnitudes were measured using circular aperture photometry, to which the aperture 
corrections derived from the PSFs were applied. The latter are the dominant source of 
error, and the nuclear magnitudes have uncertainties of 0.1 magnitudes. Table 3 also 
gives ``nuclear'' magnitudes in each of the 4  observed filters. The ({\it B$-$I}),({\it 
I$-$H}),({\it H$-$K\p}) colors of the central 2.5 kpc regions of the cool ULIGs are 
shown in Figures 6 and 7.

Most of the cool ULIG nuclei have colors consistent with a young (10-100 Myr) stellar 
population combined with hot (800K) dust emission that contributes 30--40\% of the 
K\p flux, or of a mixture of stars and absorbing dust with a total optical depth 
approaching {\it A}$_{\rm V}$ = 30--50 magnitudes. However, the more complete 
mixed stars and dust model with scattering that was discussed in Paper II indicates that 
it is difficult to achieve such reddened colors in this way, which strongly suggests that 
the ({\it H$-$K}\p) excess is actually due to hot dust emission. These colors are 
similar to the range observed by Carico et al. (1990b) in the ``LIGs'' ({\it L}$_{\rm 
IR} >$ 10$^{11}$ {\it L}$_{\sun}$). Regardless of whether they are stars with 
additional hot dust emission or just stars mixed with extinguishing dust, they 
additionally appear to be reddened by a uniform foreground dust screen of A$_V$ = 1--
5 magnitudes, which is considerably higher than any foreground screen found in the 
``warm'' ULIGs for either the nuclei or the star-forming knots. The presence of a 
greater foreground reddening screen than that in the warm ULIGs is qualitatively 
consistent with the evolution scenario in which an obscuring dust screen is blown away 
from the initial merger state.
Alone among the cool ULIGs, IRAS{\ts}22491$-$1808e appears to have optical/near-
infrared colors almost identical to that of the median ``warm'' ULIG colors. This is 
possible evidence that IRAS 22491$-$1808 harbors an AGN, although its K\p 
luminosity ({\it M}$_{\rm K^{\prime}}$ = $-$22.37) is more than an order of 
magnitude fainter than that of a QSO (Surace 1998). It is perhaps surprising that UGC 
5101 does not also have AGN-like colors. However, the {\it HST} images reveal 
structure in the vicinity of the compact {\it I}-band nucleus that cannot be resolved 
from the ground optically, and it is likely that this structure is contaminating our 
results. Results from NICMOS indicate that the point-like nucleus actually does have 
QSO-like colors (Scoville et al. 1999).

The colors of the star-forming knots are more problematic to analyze. This is because 
they are not as well determined as the nuclear colors due to their irregular shapes and 
small sizes, and because they are not detected in many cases in the near-infrared despite 
the superior resolution at long wavelengths. For the non-detections, the upper limits of 
the near-infrared luminosities of the knots constrain their colors, and hence their ages. 
Figure 8 shows the ({\it B$-$I}),({\it I$-$H}) colors of the BC95 instantaneous 
starburst used in Papers I \& II. It is apparent that for starburst colors of ({\it I$-
$H}) $<$ 1 the modeled starburst age is constrained to less than 10 Myrs. As noted in 
Paper I, with detections only at B and I and upper limits in the near-infrared, it is only 
possible to set upper limits to the knot ages; dereddening will always {\it decrease} the 
estimated ages. Many of the cool ULIGs with detectable star-forming knots (IRAS 
00091$-$0738, UGC 5101, IRAS 14348$-$1447, IRAS 20414$-$1651, IRAS 
22491$-$1808 and IRAS 23365+3604) have at least several knots whose ages cannot 
be more than 5--7 Myrs. However, several also have knots that are sufficiently red that 
their age limits can only be estimated to be less than 1 Gyr (UGC 5101, IRAS 
12112+0305, Mrk 273, IRAS 22491$-$1808) or a few hundred Myrs (IRAS 
15250+3609 and IRAS 22491$-$1808). Thus, while we can show the presence of 
young stars, we cannot easily determine a lower age limit and thus demonstrate the 
presence of intermediate age stars. Thepresence of young stars seems to be much more 
prevalent among cool ULIGs than among warm ULIGs. While this could arguably be an 
effect of greater reddening in the warm ULIGs, the results presented here run counter to 
this in that the cool ULIG nuclei seem to suffer greater foreground reddening than that 
found in the warm ULIGs.

\section{Relationship of Optical/Near-Infrared Emission to Bolometric Luminosity}

Surace \& Sanders (1999) found that in many cases the putative AGN in the warm ULIGs 
could be made to account for the high bolometric luminosities of the systems by assuming 
a QSO-like SED and extrapolating from the observed optical/near-infrared luminosities. 
It is possible to apply similar techniques to the emission in the cool ULIGs. Starburst 
models are chosen since the optical/near-infrared colors appear to be characteristic of 
young stars. The bolometric luminosity will then be determined based on both empirical 
and theoretical models. 

The theoretical SEDs are based on the BC95 models. As before, a modeled instantaneous 
starburst with a Salpeter IMF and upper and lower mass cutoffs of 0.1 and 125 \msunns 
was used. For any given age, a bolometric correction (BC) can be determined from the 
models to derive a bolometric luminosity based on the luminosity in some specific filter. 
Figure 9 shows the bolometric correction as a function of age for K-band. It is 
immediately apparent that the bolometric correction (BC) hinges critically on the age of 
the starburst. Prior to 5 Myrs, the luminosity is dominated by short-wavelength 
emission from high mass OB stars; very little of the bolometric luminosity originates in 
the late-type stars that emit strongly at long wavelengths. At 10 Myrs this changes 
radically as the most massive stars age and emit the bulk of their luminosity at 
progressively longer wavelengths, and hence the bolometric correction spans fully 6 
magnitudes depending on the age. With this model, an ultraluminous starburst could have 
M$_K < -$21.2 for a young burst, or M$_K < -$27 for an old one. Ironically, the 
change in bolometric correction with age is much less at shorter wavelengths (Figure 9 
also shows the BC as a function of age for B-band), but the uncertainty in luminosity 
caused by dust extinction at short wavelengths may offset any gain. It can be argued that 
ages of 10 Myrs and shorter for {\it all} of the knots are unlikely for several reasons. 
The cool ULIGs span a considerable range in interaction morphology. The presence of 
star-forming knots in most of these systems (as well as the more dynamically evolved, 
based on presence of single nuclei, warm ULIGs presented in Paper I), however, 
indicates that the star formation history for the knots as a whole must be similar to the 
dynamical timescale, i.e., hundreds of Myrs. Similarly, the wide range in colors seen in 
many of the knots in the ULIGs may be evidence for a spread in knot ages, although this 
may also be due to reddening.

While B-band observations are strongly affected by dust extinction (as noted earlier), 
we can constrain the maximum amount of foreground extinction on the basis that the 
starlight can only be dereddened to the colors of the bluest young starbursts. Most of the 
star-forming knots, therefore, cannot be dereddened by more than 2 magnitudes. This 
yields a fraction of the bolometric luminosity contributed by star-formation actually 
detected at B-band with a young (less than 10 Myr), dereddened starburst between 1\% 
(IRAS 00091$-$0738) and 100\% (IRAS 22491$-$1808), with a median value of 
6\%. If the starburst is older (100 Myrs), then these percentages fall by a factor of 6. 
An additional uncertainty results from the large geometric corrections to the luminosity 
discussed in Paper II. However, the scattering models indicate that it is relatively 
difficult to achieve the red colors observed in Figures 6 \& 7 via a stellar ensemble 
mixed with dust alone, and that they are more likely to be a result of hot dust emission 
and foreground extinction. If the stars are embedded in a thick dusty medium then their 
luminosities are underestimated by factors of 3--6.

Results for the K-band are least affected by uncertainties in dust extinction, and hence 
should give the best luminosity estimate. We have examined those cases where star 
formation can be morphologically separated from the underlying host galaxy stellar 
population, i.e. those systems that show evidence for star-forming knots. The total K-
band flux was dereddened by the amount indicated by the nuclear optical/near-infrared 
colors and then the portion attributable to starlight (typically 50--70\%, again 
determined by the colors) was separated out. These extinction estimates are likely to be 
overestimates, given the high extinctions derived from the nuclear colors and the 
maximum extinctions derived from just the knot optical colors. As discussed above, the 
BC95 models cannot readily constrain the bolometric luminosity based on the K-band 
luminosity due to the enormous age dependence of the BC. If the star-forming knots are 
very young, (less than 1 Myr), then every cool ULIG with detectable knots at K\p could 
conceivably derive its entire bolometric luminosity from the star-forming knots alone. 
The upper limits for the knots not detected at K\p can only constrain the bolometric 
luminosity under this assumption of extremely young stellar age to being just under the 
measured ULIG bolometric luminosities. If the knots are more than 10 Myrs in age, then 
it is likely that none of the ULIGs could have contributions to their bolometric 
luminosities by star-forming knots much above 50\%, and in most cases it would be 
less than 10\%.

The empirical model is based on the bolometric correction from K-band to {\it 
L}$_{\rm ir}$ found in the LIGs (Carico et al. 1990; equation 7 Paper II):

\begin{equation}
{\rm log} L_{\rm ir}=-{{M_{\rm K^{\prime}}-6.45}\over{2.63}}
\end{equation}

\noindent Assuming that nearly all ($\approx$95\% )of the bolometric luminosity is 
emitted in the far-infrared (Sanders et al. 1988a), this is equivalent to a bolometric 
correction of 0-0.3 for M$_{K^{\prime}}$=$-$20 to $-$25, or roughly equivalent to 
the modeled value for a starburst 10 Myrs old (i.e. a BC of 0--0.3 will convert between 
M$_{K^{\prime}}$ to M$_{\rm bol}$ in LIGs). Using equation 1, the derived total 
bolometric luminosity for the detected star-forming knots ranges from 10$^{10}$ 
\lsun to 10$^{11.4}$ \lsunns . The typical ULIG which has star-forming knots detected 
in {\it any} band has a contribution to the bolometric luminosity of not more than 
10$^{10.5}$ \lsunns . This falls short of the average cool ULIG bolometric luminosity 
by a factor of 50. It thus appears that {\it nothing} detected optically or in the near-
infrared in the cool ULIGs is capable of generating the high bolometric luminosity 
assuming the kinds of SEDs we have used here. Note that this result does not preclude the 
existence of an ultraluminous starburst or AGN, since ultimately something must 
provide the known bolometric luminosity. Rather, this result implies that no such 
object is directly observable in the optical or near-infrared. Whatever the power 
source, it must be more highly obscured in the cool ULIGs than in the warm, a result 
supported by the estimated extinctions derived from the optical/near-infrared colors.

\section{Conclusions}

We have presented high spatial resolution images of a sample of ``cool'' ULIGs. We find 
that:

\noindent 1. All of the systems are major mergers, as manifested by prominent tails and 
other extended tidal structure.

\noindent 2. A large fraction (at least 43\% and as high as 72\%) have resolvable 
double galactic nuclei. Their projected separations span the range 2.5--25 kpc. Double 
nuclei could have been detected with separations as small as 600 pc. The lack of small-
separation (0.6--2.5 kpc) systems may support earlier similar findings that indicate 
that the time for final coalescence of the nuclei is comparatively brief.

\noindent 3. Most of the ``cool'' ULIGs have evidence for compact star-forming knots, 
with the exception of the systems with the widest separations. This may indicate that 
clustered formation begins in earnest only near the end stages of the merger, just before 
nuclear coalescence, although this result may also partly be due to the limited ground-
based resolution.

\noindent 4. The nuclear colors of most of the cool ULIGs appear similar to that of a 
mixture of stars and extinguishing dust with a total optical depth of {\it A}$_{\rm V}$ = 
30--50 magnitudes, or of young stars with a modest amount of the {\it K}\p emission 
(30\%) originating in hot (800 K) dust. This hot dust emission is then further 
extinguished by a uniform dust screen {\it A}$_{\rm V}$ = 1--5 magnitudes thick. 
Unlike the ``warm'' ULIGs, the optical/near-infrared emission from the nuclear regions 
of the cool sample is probably stellar in nature.

\noindent 5. The dereddened {\it H}-band luminosities of the cool ULIG host galaxies lie 
in the range 0.9--2.5 {\it L}$^*$, and are thus essentially identical to those of the 
warm ULIGs and are consistent with their apparent merger origin. There are, however, 
no systems with {\it L}$_{\rm H} > $ 3 {\it L}$^*$, unlike 25\% of the warm ULIGs.

\noindent 6. Very few of the star-forming knots are detected in the near-infrared, nor 
are any new knots (much like the ``warm'' ULIGs). Constraints imposed by the limits of 
their ({\it I-H}) colors imply very young ages ($<$ 5--7 Myrs) for many of these 
knots. They cannot be extinguished by more than a very mild foreground reddening 
screen ($<$ {\it A}$_{\rm V}$ = 2 magnitudes), and any additional knots must be very 
deeply embedded.

\noindent 7. As in the ``warm'' ULIGs, it appears that in most cases the star-forming 
knots are insufficiently luminous to be the source of the high bolometric luminosity, 
although in some cases they may provide a significant fraction. Although the constraints 
are found to be very model dependent, using assumptions similar to those used in Paper 
II, the observed optical/near-infrared emission observable in the knots provides 
typically only about 2\% of the high bolometric luminosity, ranging from less than 1\% 
to about 20\%. This is very similar to the results found for the warm ULIGs. It appears 
unlikely that anything seen in the optical or near-infrared is related to the high 
bolometric luminosity, unless it has an SED much more biased towards the far-infrared.

\acknowledgments

We thank the creators of the tip/tilt system and the instruments, Kevin Jim and Gerry 
Luppino.
We thank John Dvorak, Chris Stewart, and Rob Whitely for operating the telescope, and 
Andrew Pickles for helping debug numerous telescope-related problems.
We thank Josh Barnes for his useful comments on early drafts of this text and Atiya 
Hakeem for proofreading it. We also thank Catherine Ishida and Alan Stockton for helping 
to supply observing time to complete these observations. 
We thank an anonymous referee whose comments helped improve the presentation of this paper. D.B.S was supported in part by 
JPL contract no. 961566 and J.A.S. was supported in part by NASA grant NAG5-3370.

\appendix

\section{Notes on Individual Objects}

{\it IRAS{\ts}00091$-$0738} --- a  system very similar in appearance to 
IRAS{\ts}12112+0305 with an extremely complex nuclear core 3.5 kpc in diameter 
bifurcated N-S by dust lanes. A thick, perhaps edge-on tail extends 19 kpc (projected 
distance) to the south, while another plume-like tail extends 20 kpc to the north. At 
near-infrared wavelengths the core becomes a single source 0.7\arcsec \ in diameter, 
with a tidal tail to the north. This connects to and appears to be the base of the optical 
tidal tail which loops around back towards the south, and appears similar to the 
structure seen in the core of IRAS 12112+0305. Identified as having HII spectra by 
Veilleux et al. (1998).

{\it IRAS{\ts}01199$-$2307} --- double nucleus system separated by 25 kpc, the 
largest separation of any ULIG in this sample. Both nuclei are ellipsoidal in appearance 
and remain similar to spiral bulges; Veilleux et al. (1998) identifies these as having 
HII spectra. The NE nucleus is surprisingly faint in the near-infrared compared to the 
SW; additionally, at K\p it takes on the appearance of two lobes, the SW of which seems 
to correspond to the optical nucleus, and the NE to the arm extending from it. The 
western tail is 48 kpc in total length, with an apparent projected linear length of 40 
kpc. The fainter eastern tail is 30 kpc long.

{\it IRAS{\ts}03521+0028} --- double nucleus system separated by 4.3 kpc. Short, 
stubby tails extend 9 kpc from each nucleus in an E-W direction. No additional star-
forming knots or other structure are seen. This galaxy has LINER optical spectra 
(Veilleux et al. 1997).

{\it UGC{\ts}5101} - a nearby system from the BGS, identified as a Seyfert 1.5 by 
Sanders et al. (1988a). A linear tail stretches 38 kpc to the west from the nucleus. A 
second tail runs clockwise from a position angle of -90 to form a nearly complete circle 
225 degrees around with a radius of 17 kpc and a total length of 65 kpc. The optical 
morphology seems to suggest that the two sets of tails are actually similar features lying 
in planes perpendicular to each other. The {\it HST} images clearly reveal a set of spiral 
dust lanes to the north of the nucleus; the nucleus itself is dominated by a small (200 
pc) size emission region. This spiral structure, along with that in the core of Mrk 273, 
may resemble that seen in the warm ULIG Mrk 231 (Paper I), only rotated out of the 
plane of the sky.

{\it IRAS{\ts}12112+0305} --- this galaxy appears to consist of a double-lobed core 
similar to that seen in IRAS{\ts}00091-0738, with a bright tidal tail extending 18 kpc 
to the north and another arc 30 kpc long loops along the south. The orientation of the 
tails suggests that the northern tail is parallel to the line of sight, while the southern 
lies tilted by 45 \degree . The southern arc has a blue condensation halfway along its 
length. A red star-like object is located 4 kpc SW of the core. Seen by Carico et al. 
(1990), this is unlikely to be a supernova since it has not faded noticeably in the 
intervening 7 years. It is not a foreground star as recent near-infrared spectroscopy 
(Surace, in prep) indicates that it is at the same redshift as the nucleus to its north --- 
if it is a starburst knot or AGN then its total lack of U\p emission indicates a line of sight 
extinction of at least {\it A}$_{\rm v}$ = 3 magnitudes (Surace \& Sanders 1999b, in 
press). Kim (1995) identifies this ULIG as having a LINER spectrum.

{\it Mrk{\ts}273}  --- a very narrow ($\approx$ 2.5kpc) linear tidal tail extends 41 
kpc south of the nucleus. Two diffuse plumes extend north and northeast 40 kpc each. The 
new deep images suggest that these plumes actually connect to the NE, making a complete 
ring nearly 100 kpc in circumference. In this respect it may be very similar to 
UGC{\ts}5101, which it strongly resembles. The {\it HST} image of the galaxy core at I-
band clearly shows a pattern of dust lanes running along the long axis of the nucleus, 
closely resembling those seen in edge-on spiral galaxies. The nucleus is dominated by a 
small, yet extended emission region. Khachikian \& Weedman (1974) identify this as a 
Seyfert 2. Several authors have shown evidence for an apparent double nucleus in Mrk 
273 (Majewski et al. 1993 Condon et al. 1991), but more recent adaptive optics 
imaging at K-band indicates that this ``nucleus'' is more likely to be a luminous star-
forming region (Knapen et al. 1997).

{\it IRAS{\ts}14348$-$1447} --- two nuclei separated by 5.3 kpc. A plume extends 
north 20 kpc from the NE nucleus. A second plume stretches 17 kpc to the SW from the 
SW nucleus, where it merges with a fan-shaped plateau of emission extending from the 
NE to the SW of the galaxy. At least a dozen star-forming knots are seen in the {\it HST} 
image. The circular ring of knots surrounding the SW nucleus are well detected in our B 
\& I images, as are the knots in the base of the northern tail. We fail to detect these (or 
any other) knots in either near-infrared filter. Although Nakajima et al. (1991) 
claimed detection of a broad-line component in the SW nucleus, Veilleux et al. (1997) 
fail to confirm this and designate this galaxy as a LINER.

{\it IRAS{\ts}15250+3609} --- an apparently single nucleus system, this galaxy has 
at least three other galaxies nearby to the north, south and west.  However, it is unclear 
if these galaxies are physically related to the merger system. A tidal feature appears to 
emerge from the SW side of the nucleus and loops around on the eastern side to create a 
closed ring 27 kpc in diameter and 80 kpc long. Another, apparently shorter arm 
extends from the northeast of the nucleus. Near-infrared imaging reveals several 
additional knots of star formation near the nucleus. Veilleux et al. (1995) classify this 
as a LINER.

{\it Arp 220} --- the closest and hence most well-studied of all ULIGs. A short tail 
extends approximately 30 kpc to the northwest (Sanders et al. 1988a; Hibbard 1995). 
Graham et al. (1990) have shown evidence for a double ``nucleus'' in Arp 220. Recent 
NICMOS imaging of the same galaxy has shown two nuclei separated by 360 pc, along 
with considerable star formation (Scoville et al. 1998). Because this ULIG appears be 
radically different from most others in terms of its high degree of variable dust 
obscuration, the registration between our ground-based optical data and the NICMOS data 
is highly uncertain. Therefore, the 2.5 kpc aperture photometry should be regarded with 
caution; we have assumed that the bright near-infrared peaks are spatially coincident 
with the optical dust lane. Kim (1995) classifies this as a LINER galaxy. The reader is 
directed towards a wealth of literature on this object (Sanders et al. 1988a; Graham et 
al. 1990; Skinner et al. 1997; Scoville et al. 1998).

{\it IRAS{\ts}20414$-$1651} --- this complex system is somewhat different from 
the other ULIGs. It has a horseshoe-shaped main body with some sort of extended 
structure ``corkscrewing'' 17 kpc to the south, which then bends west and meets with a 
very blue stellar condensation. These condensations may be superimposed background 
objects, or newly formed high density regions in the tails themselves. Kim (1995) 
identifies this as having HII spectra.

{\it IRAS{\ts}22206$-$2715} --- two nuclei separated by 9.2 kpc. The northern 
nucleus is circular in shape, perhaps suggesting that it is viewed face on. The southern 
nucleus is elongated and bar-shaped, suggesting a spiral galaxy inclined by roughly 
60\degree , an idea supported by the zig-zag nature of the tidal tail emanating from it. 
Each tail is approximately 20 kpc in total length. Like many of the other galaxies here, 
this seems to be a collision between two galaxies with high relative inclination, 
resulting in one broad, almost circular tidal tail, and one that is seen edge-on, or nearly 
so. Veilleux et al. (1998) finds this galaxy to have an HII spectrum.

{\it IRAS{\ts}22491$-$1808} --- the most extreme example of clustered star-
formation in the sample. At optical wavelengths the knots of star-formation create so 
much confusion as to preclude identification of the galaxy nuclei. Only in the near-
infrared do the two nuclei, separated by 2.5 kpc, stand out. The system has two high 
surface brightness tails. The first extends 12.5 kpc NE from the main body of the galaxy 
and is essentially featureless. The second extends NW 16 kpc from the main body, but 
ends in a complex, face-on circular loop 10 kpc in diameter. This tail has two very red 
clumps of star formation at its base, and the circular disk at the end of the tail has many 
blue knots of star formation. The {\it HST} data seem to suggest that the NE tail may also 
terminate in a somewhat fainter version of this disk. This galaxy has an HII spectrum 
(Sanders et al. 1988a; Veilleux et al. 1995).

{\it IRAS{\ts}23233+0946} --- two nuclei separated by 8.5 kpc. A tidal tail 19 kpc in 
linear distance and apparently 28 kpc in total length extends to the SE. Both nuclei have 
colors consistent with an old stellar population. Veilleux et al. (1998) classify this as a 
LINER.

{\it IRAS{\ts}23365+3604} --- this galaxy has a single, point-like nucleus embedded 
in a face-on disk 20 kpc in diameter. Four large star-forming knots are contained in 
this disk, and are detected even in the near-infrared. The disk itself has a twisting spiral 
structure. A faint linear tail extends north for 60 kpc from the nucleus. Another, higher 
surface brightness tail extends 20 kpc due south from the nucleus. This tail has an odd 
projection halfway along its length --- a short (few kpc) tail jutting west from its 
side, not unlike the feature seen in the linear tail of Mrk 273. Veilleux et al. (1995) 
classify this as a LINER.

\clearpage

\figcaption{Optical (B,I) and near-infrared (H,K\p) images of the ``cool'' ULIGs 
showing their large-scale structure. Northeast is at top left. All of the images for a given 
object except for Arp 220 and UGC 5101 are at the same scale, with ticks 2\arcsec \  
apart and major ticks every 10\arcsec . The physical scale bar on each of the B-band 
images is 10 kpc in length. Contours are logarithmic and illustrate the high surface 
brightness morphology.}

\figcaption{Near-truecolor images of the ``cool'' ULIGs constructed from the B \& I-
band data. The galaxy SEDs have been linearly interpolated from the {\it B} and {\it I} 
data; the color balance is not absolute.}

\figcaption{{\it HST}/WFPC2 archival F814W (I-band) images of ULIGs from the 
``cool'' ULIG sample. Each tick mark is 1\arcsec , and the physical scale-bar is 2.5 kpc. 
Orientation is with northeast at upper-left.}

\figcaption{An illustration of the possible ULIG merger progenitor geometry: a plunging 
collision between two disks highly inclined to each other. At upper left is a schematic 
representation of such a collision. Two disks are inclined at nearly right angles to each 
other and meet in a slightly off-center (non-zero impact parameter) plunging 
encounter. Upper right and bottom left show two orientations of the same 1.4 Gyr-old 
collision between two spiral galaxies, one of whose disks is in the orbital plane, and the 
other inclined to it by 71 \degree (Barnes 1992, Barnes \& Hernquist 1996). This type 
of encounter raises a nearly circular, off-center tail structure in the lower left galaxy 
as a result of the plunging orbit (Toomre 1978). At bottom right is an I-band image of 
UGC 5101 for comparison.}

\figcaption{Selected images of high spatial resolution structure in the nuclear regions of 
the cool ULIGs. All of the optical data except IRAS 12112+0305 and Mrk 273 has been 
deconvolved with the Richardson-Lucy maximum likelihood algorithm. The physical 
scale-bar is 2.5 kpc in length. Tick marks are every 2\arcsec \ with major ticks every 
10\arcsec .}

\figcaption{({\it B$-$I},{\it I$-$H},{\it H$-$K\p}) color cube illustrating the colors 
of the ``cool'' ULIG nuclear (2.5 kpc diameter) regions. For clarity, the ULIGs have been 
marked with numbers: (1) IRAS 00091$-$0738 (2) IRAS 01199$-$2307w (3,4) 
IRAS 03521+0028e,w (5) UGC 5101 (6,7) IRAS 12112+0305n,s (8,9) IRAS 
14348$-$1447e,w (10) IRAS 15250+3609 (11) IRAS 20414$-$1651 (12,13) 
IRAS 22206$-$2715 (14,15) IRAS 22491$-$1448 (16,17) IRAS 23233+0946 
and (18) IRAS 23365+3604. The cube is rotated to be orthogonal to the reddening 
vector, which is depicted by the closed boxes and represents line-of-sight extinction, 
i.e. a simple foreground dust screen, in 
units of {\it A}$_{\rm V}$ = 1 magnitude. It is derived from Rieke \& Lebofsky 
(1985). 
The median colors of the ``warm'' ULIGs are marked with a large circle. 
The open circles are the colors of an instantaneous starburst with a Salpeter IMF and 
mass range 0.1--125\msun and aging from 0 to 15 Gyrs, based on the spectral 
synthesis models of Bruzual \& Charlot (1993; an updated version called BC95 is used 
here).
The large dotted open circle is a synthetic QSO spectrum based on multiwavelength 
observations of Palomar-Green QSOs and is discussed in detail in Paper II. It is 
indicative of the colors of typical optically selected quasars.
The joined, open circles show the effects of adding free-free emission with a 20,000 K 
electron temperature in increments of 
20\% of the total flux at K\p to the starburst.
The two sets of filled, joined circles illustrate emission from 800 K dust contributing in 
increments of 10\% to the total flux at K\p.
Finally, emission from uniformly mixed stars and dust, in units of {\it A}$_{\rm 
V}$=10, 30, 
and 50 magnitudes, are shown by the $\times$ symbol.
One $\sigma$ error bars are shown at upper right. Note that the line-of-sight dust 
extinction and thermal dust emission curves are nearly 
orthogonal.}

\figcaption{Same as Figure 6, but rotated as to be parallel to the reddening vector. Error 
bars are at upper left.}

\figcaption{({\it B$-$I}),({\it I$-$H}) color-color diagram for an instantaneous 
starburst having a Salpeter IMF and a mass range from 0.1--125 \msunns. Also shown 
are the colors of the star-forming knots given in Table 4.}

\figcaption{Bolometric corrections for {\it K}-band (circles) and {\it B}-band 
(triangles) for an instantaneous starburst with a Salpeter IMF ranging from 0.1 to 125 
M$_{\sun}$ (BC95).}

\clearpage
\vspace*{4.0truein}
Figures 1-5 are images and are available in JPEG format either from this preprint archive or from http://humu.ipac.caltech.edu/preprints/cool.

\setcounter{figure}{5}
\begin{figure}
\epsscale{0.9}
\plotone{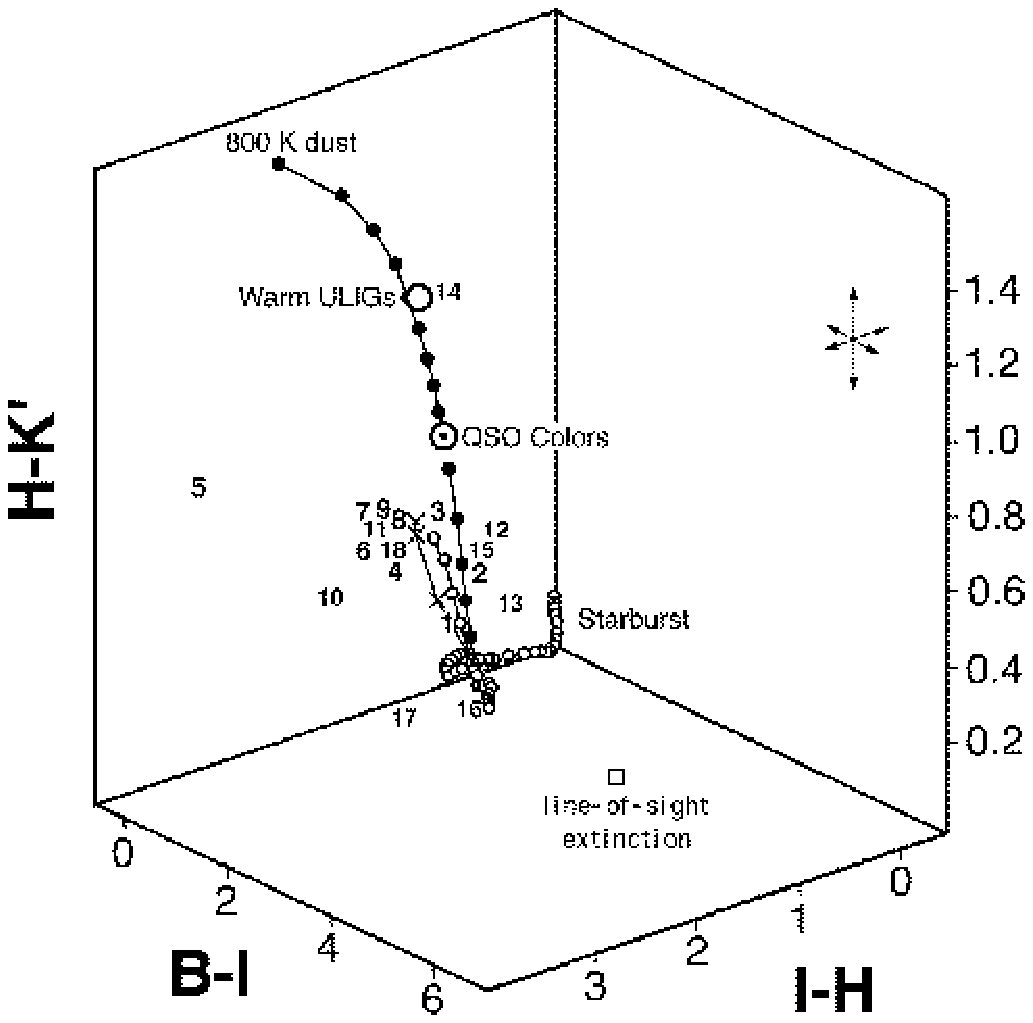}
\caption{}
\end{figure}
\clearpage

\begin{figure}
\epsscale{0.9}
\plotone{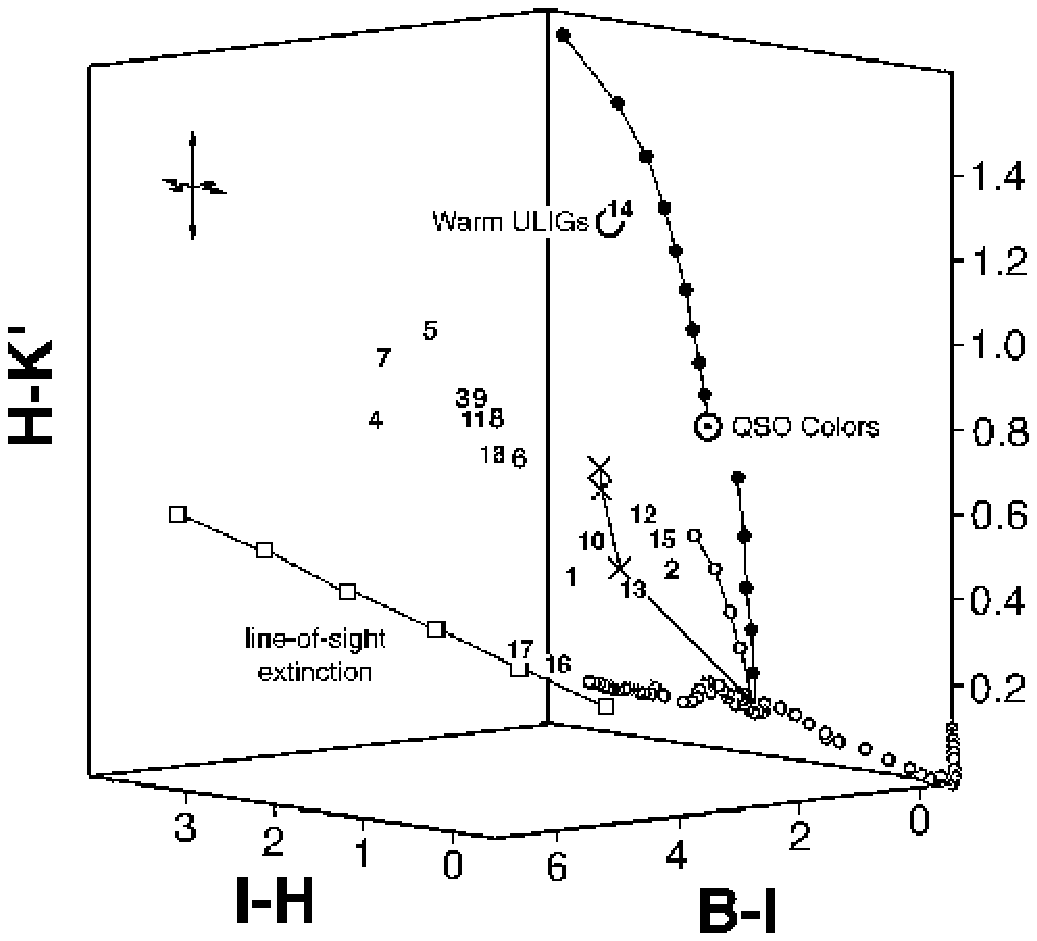}
\caption{}
\end{figure}
\clearpage

\begin{figure}
\epsscale{0.9}
\plotone{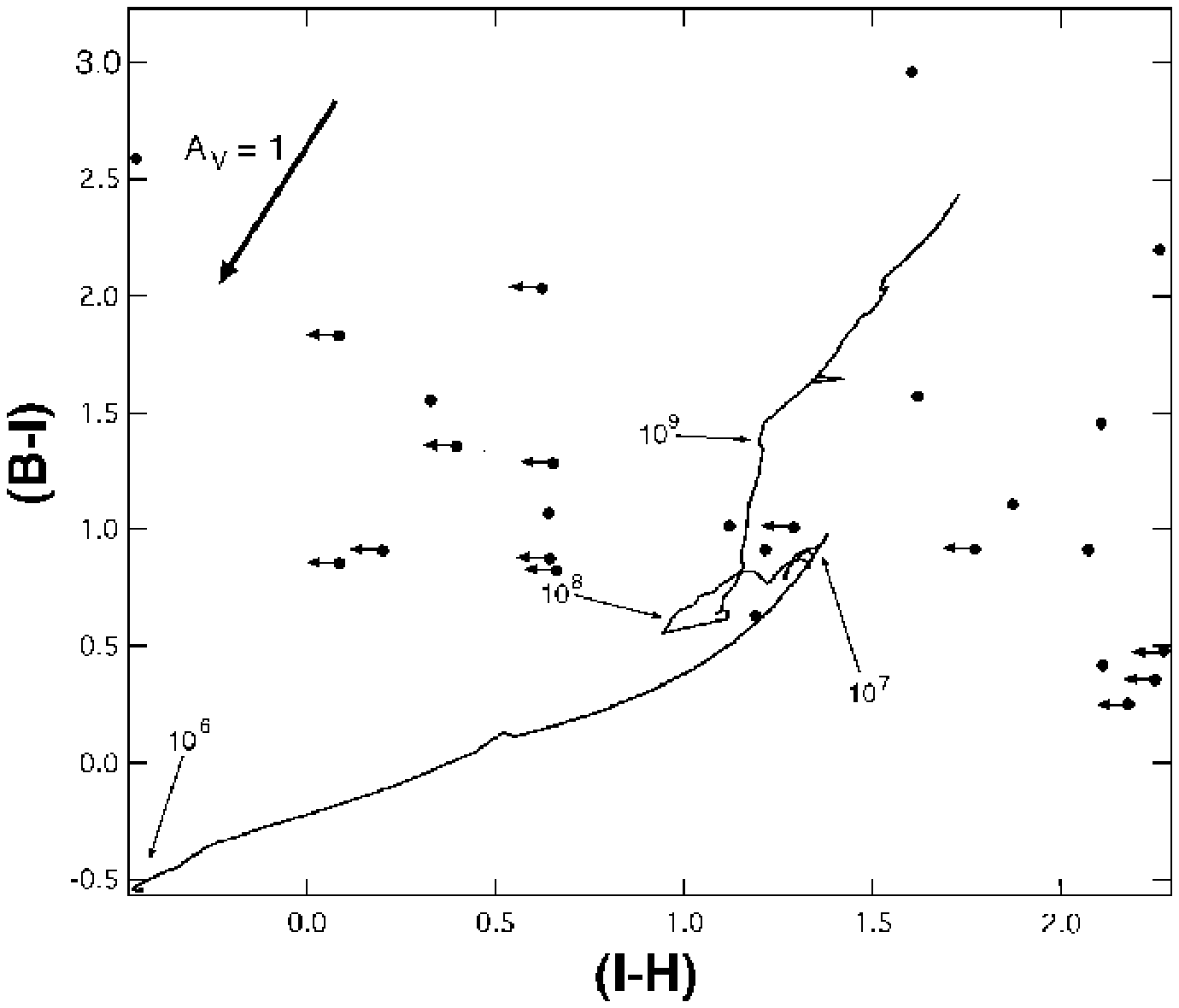}
\caption{}
\end{figure}
\clearpage

\begin{figure}
\epsscale{0.9}
\plotone{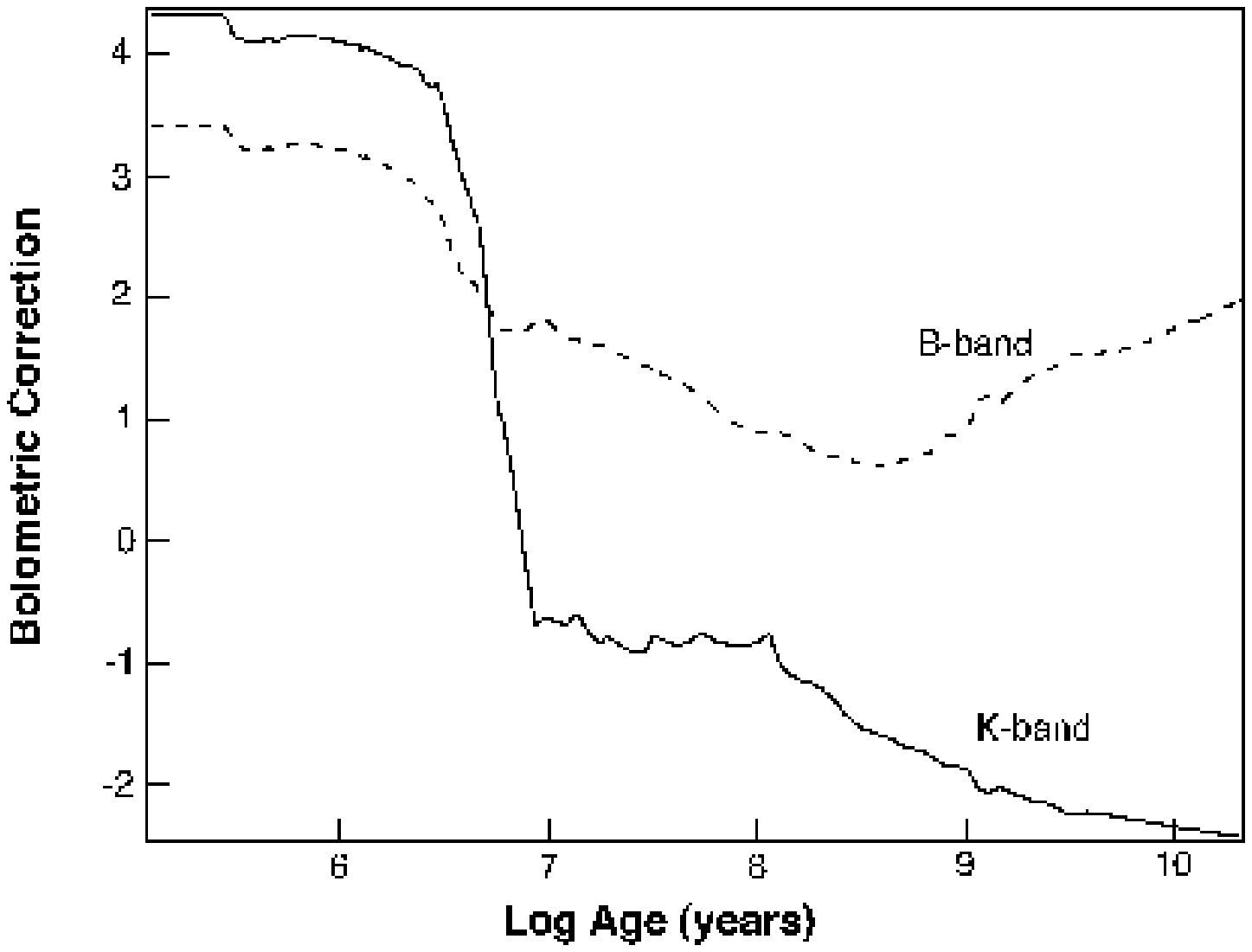}
\caption{}
\end{figure}
\clearpage

\clearpage

{\small
\begin{deluxetable}{lrrrrrrrrr}
\tablewidth{0truein}
\tablecaption{Cool ULIG Sample \& Observation Details}
\tablehead{
\colhead{Name} &
\colhead{RA$^1$} &
\colhead{DEC} &
\colhead{z} &
\colhead{{\it L}$_{\rm ir}$} &
\multicolumn{1}{c}{Inst.$^2$} &
\multicolumn{4}{c}{Exposure Time (sec)} \\[0.2ex] 
\colhead{} &
\multicolumn{2}{c}{(J2000.0)} &
\colhead{} &
\colhead{{\it L}$_{\sun}$} &
\multicolumn{1}{c}{} &
\multicolumn{1}{c}{B} &
\multicolumn{1}{c}{I} &
\multicolumn{1}{c}{H} &
\multicolumn{1}{c}{K\p}
}
\startdata
IRAS{\ts}00091$-$0738 & 00:11:43.2 & $-$07:22:07.8 & 0.119 & 12.17 & QT & 2520 & 1920 & 2520 & 1800 \nl
IRAS{\ts}01199$-$2307 & 01:22:21.4 & $-$22:51:59.5 & 0.156 & 12.23 & QT & 2880 & 2280 & 2400 & 2160 \nl
IRAS{\ts}03521+0028 & 03:54:42.1 &    00:37:05.9 & 0.152 & 12.44 & QT & 3600 & 2280 & 2400 & 3360 \nl
UGC{\ts}5101        & 09:35:51.7 &    61:21:11.3 & 0.039 & 11.99 & WQT & 1620 & 2160 & 2160 & 2040  \nl
IRAS{\ts}12112+0305 & 12:13:46.1 &    02:48:41.4 & 0.072 & 12.24 & WQT & 1440 & 1860 & 1920 & 1440 \nl
Mrk{\ts}273         & 13:44:42.1 &    55:53:12.7 & 0.038 & 12.13 & W\phantom{Q}T & 1200 & 2340 & \nodata & \nodata \nl
IRAS{\ts}14348$-$1447 & 14:37:38.7 & $-$15:00:22.8 & 0.082 & 12.24 & WQH & 720 & 720 & 1440 & 1680 \nl
IRAS{\ts}15250+3609 &  15:26:59.4 &    35:58:37.5 & 0.055 & 11.99 & QH & 1200 & 1680 & 2160 & 2160 \nl
Arp 220             & 15:34:57.3 &    23:30:11.9 & 0.018 & 12.17 & NT & 1080 & 1440 & \nodata & \nodata \nl
IRAS{\ts}20414$-$1651 & 20:44:17.4 & $-$16:40:13.7 & 0.086 & 12.12 & WQH & 1440 & 1560 & 1680 & 1800 \nl
IRAS{\ts}22206$-$2715 & 22:23:28.8 & $-$27:00:03.3 & 0.131 & 12.15 & QT & 480 & 1440 & 2400 & 2520 \nl
IRAS{\ts}22491$-$1808 & 22:51:49.3 & $-$17:52:23.5 & 0.078 & 12.08 & WQT & 2520 & 1440 & 2760 & 2520 \nl
IRAS{\ts}23233+0946 & 23:25:55.6 &    10:02:45.1 & 0.128 & 12.02 & QT & 2160 & 1800 & 2160 & 2160 \nl
IRAS{\ts}23365+3604 & 23:39:01.3 &    36:21:09.8 & 0.064 & 12.10 & QH & 240 & 240 & 1200 & 1560 \nl
\enddata
\tablenotetext{}{$^1$ Positions derived from Kim (1995), redshifts and {\it L}$_{\rm ir}$ taken from Kim (1995) and Sanders et al. (1988a)}
\tablenotetext{}{$^2$ Q = UH2.2m f/31 QUIRC, T = UH2.2m f/31 Tektronix 2048, H = UH2.2m f/31 Orbit reimaged at f/10 through HARIS spectrograph, 
W = {\it HST}/WFPC2, N={\it HST}/NICMOS}    
\end{deluxetable}
}



{\small
\begin{deluxetable}{llcrccr}
\tablewidth{0truein}
\tablecaption{Cool ULIG Morphology}
\tablehead{
\colhead{Name} &
\multicolumn{1}{c}{Nuclei$^1$} &
\multicolumn{1}{c}{Nuclear Sep.} &
\multicolumn{1}{c}{Tails$^2$} &
\multicolumn{1}{c}{Optical Knots$^3$} &
\multicolumn{1}{c}{Near-IR Knots} &
\multicolumn{1}{c}{Spectral Class} \\ [0.2ex]
\multicolumn{2}{c}{ }&
\multicolumn{1}{c}{kpc} &
\multicolumn{4}{c}{ }
}
\startdata
IRAS{\ts}00091$-$0738 & SB & \nodata & L & Y & N & HII \nl
IRAS{\ts}01199$-$2307 & D & 25.0\phantom{0} & C & N & N & HII \nl
IRAS{\ts}03521+0028   & D & 4.3 & C & N & N & LINER  \nl
UGC{\ts}5101        & SB & \nodata & LC & Y & Y & Sy1.5  \nl
IRAS{\ts}12112+0305 & SBD? & 4 & C & Y & Y & LINER \nl
Mrk{\ts}273         & SBD? & \nodata & LC & Y & Y & Sy2 \nl
IRAS{\ts}14348$-$1447 & D & 5.3 & L & Y & Y & LINER \nl
IRAS{\ts}15250+3609 & SC & \nodata & C & Y & Y & LINER \nl
Arp 220             & SBD? & \phantom{0}0.4$^4$ & L & Y & Y & LINER \nl
IRAS{\ts}20414$-$1651 & D? & \phantom{?}2.6? & C & Y & N & HII \nl
IRAS{\ts}22206$-$2715 & D & 9.2 & L  & N & N & HII \nl
IRAS{\ts}22491$-$1808 & D & 2.5 & LC & Y & Y & HII \nl
IRAS{\ts}23233+0946 & D & 8.5 & C & N & N & LINER \nl
IRAS{\ts}23365+3604 & SC & \nodata & LC & Y & Y & LINER \nl
\enddata
\tablenotetext{}{$^1$ D=Double, D?=Possible Double, SC=Single Compact, SB=Single,Bifurcated }
\tablenotetext{}{$^2$ L=Linear, C=Circular or semi-circular}  
\tablenotetext{}{$^3$ Y=Yes, N=No}  
\tablenotetext{}{$^4$ taken from Scoville et al. (1998)}
\end{deluxetable}
}



{\small
\begin{deluxetable}{lcccccccc}
\tablewidth{0truein}
\tablecaption{Cool ULIG Total and Nuclear Photometry}
\tablehead{
\colhead{Name} &
\multicolumn{2}{c}{{\it m}$_{\rm B}$} &
\multicolumn{2}{c}{{\it m}$_{\rm I}$} &
\multicolumn{2}{c}{{\it m}$_{\rm H}$} &
\multicolumn{2}{c}{{\it m}$_{\rm K^{\prime}}$} \\ [0.2ex] 
\colhead{} &
\colhead{Total} &
\colhead{Nuclear$^1$} &
\colhead{Total} &
\colhead{Nuclear} &
\colhead{Total} &
\colhead{Nuclear} &
\colhead{Total} &
\colhead{Nuclear} 
}
\startdata
IRAS{\ts}00091$-$0738 & 18.37 & 19.4 & 16.36 & 17.3 & 14.81 & 15.3 & 14.31 & 14.9 \nl
IRAS{\ts}01199$-$2307 & 17.82 & 19.5 & 16.69 & 18.0 & 15.11 & 16.2 & 14.85 & 16.2 \nl
IRAS{\ts}03521+0028 & 20.50 & 22.9 & 17.49 & 19.3 & 14.91 & 16.1 & 14.11 & 15.3 \nl
UGC{\ts}5101        & 14.91 & 17.9 & 14.60 & 16.5 & 11.53 & 12.4 & 10.83 & 11.5 \nl
IRAS{\ts}12112+0305n & 16.97 & 20.0 & 15.05 & 18.0 & 12.88 & 15.3 & 12.76 & 14.6 \nl
IRAS{\ts}12112+0305s & & 21.6 & & 18.4 & & 15.0 & & 14.1 \nl
Mrk{\ts}273         & 14.69 & 17.1 & 13.16 & 15.3 & \nodata & \nodata  & \nodata & \nodata \nl
IRAS{\ts}14348$-$1447e & 16.51 & 20.6 & 14.58 & 18.1 & 12.92 & 15.5 & 12.19 & 14.7  \nl
IRAS{\ts}14348$-$1447w & & 20.1 & & 17.6 & & 14.9 & & 14.0 \nl
IRAS{\ts}15250+3609 & 16.17 & 17.5 & 15.19 & 16.6 & 12.98 & 14.0 & 12.77 & 13.6 \nl
Arp 220             & 14.00 & 17.6 & 12.25 & 14.7 & 10.93 & 11.6 & 10.62 & 10.9 \nl
IRAS{\ts}20414$-$1651  & 17.94 & 20.2 & 16.11 & 17.7 & 14.02 & 14.8 & 13.49 & 14.1 \nl 
IRAS{\ts}22206$-$2715e & 17.26 & 18.7 & 15.78 & 16.9 & 14.01 & 15.3 & 13.45 & 14.7 \nl
IRAS{\ts}22206$-$2715w & & 19.3 & & 17.2 & & 15.9 & & 15.5 \nl
IRAS{\ts}22491$-$1808e & 16.44 & 18.6 & 15.00 & 17.1 & 13.26 & 15.7 & 12.97 & 15.2 \nl
IRAS{\ts}22491$-$1808w & & 19.9 & & 18.2 & & 16.3 & & 15.1 \nl
IRAS{\ts}23233+0946e & 18.12 & 20.9 & 15.80 & 18.3 & 14.16 & 16.4 & 13.80 & 16.2 \nl
IRAS{\ts}23233+0946w & & 19.7 & & 17.4 & & 14.9 & & 14.7 \nl
IRAS{\ts}23365+3604 & 16.17 & 19.5 & 14.32 & 17.1 & 12.68 & 14.5 & 12.19 & 13.8 \nl
\tablenotetext{}{$^1$ central region 2.5 kpc in diameter. Total magnitudes
are listed only under the first nucleus, but refer to the entire system.
Ellipses indicate missing data.}
\tablenotetext{}{Total magnitudes are uncertain to $\pm$0.07 magnitudes. Nuclear magnitude uncertainties are $\pm$0.1 magnitudes.}
\enddata
\end{deluxetable}
}



{\small
\begin{deluxetable}{lrrrrrr}
\tablewidth{0truein}
\tablecaption{Star-Forming Knot Photometry}
\tablehead{
\colhead{Number} & 
\colhead{$\Delta$RA$^1$} &
\colhead{$\Delta$DEC} &
\colhead{{\it m}$_{\rm B}$} &
\colhead{{\it m}$_{\rm I}$} &
\colhead{{\it m}$_{\rm H}$} &
\colhead{{\it m}$_{\rm K^{\prime}}$}
}
\startdata
\cutinhead{IRAS 0009$-$0738}
 1 & 1.1 & 0.3 & 21.23 & 19.11 & $>$ 18.50 & $>$ 18.61 \nl
\cutinhead{UGC 5101}
 1 & 1.9 & 0.0 & 18.79 & 18.00 & $>$ 17.92 & $>$ 17.93 \nl
 2 & $-$5.6 & $-$0.3 & 20.24 & 20.04 & $>$ 17.63 & $>$ 18.66 \nl
 3 & $-$4.6 & 0.7 & 20.03 & 19.85 & $>$ 17.63 & $>$ 18.66 \nl
 4 & $-$3.8 & 2.9 & 20.42 & 20.03 & $>$ 17.63 & $>$ 18.66 \nl
 5 & 9.4 & 1.4 & 20.61 & 20.27 & $>$ 17.90 & $>$ 19.00 \nl
 6 & 6.3 & 10.0 & 21.63 & 20.15 & 18.07 & 18.03 \nl 
\cutinhead{IRAS 12112+0305}
 1 & $-$2.8 & 0.3 & 20.32 & 18.22 & 15.92 & 15.53 \nl
 2 & $-$2.7 & $-$2.8 & 21.62 & 18.37 & 15.09 & 14.33 \nl
 3 & $-$2.8 & $-$7.4 & 19.90 & 18.31 & 16.69 & 16.32 \nl
\cutinhead{Mrk 273}
 1 & $-$0.7 & $-$3.64 & 17.61 & 16.09 & \nodata & \nodata \nl
\cutinhead{IRAS 14348$-$1447}
 1 & $-$1.1 & 1.3 & 19.31 & 18.42 & $>$ 17.75 & $>$ 17.17 \nl
 2 & $-$1.3 & $-$1.4 & 19.69 & 18.42 & $>$ 17.75 & $>$ 17.17 \nl
 3 & 3.2 & 4.1 & 19.12 & 18.25 & $>$ 18.00 & $>$ 17.03 \nl
\cutinhead{IRAS 15250+3609}
 1 & 0.5 & $-$0.4 & \nodata & \nodata & 14.39 & 13.96 \nl
 2 & 1.5 & 1.4 & 20.50 & 17.86 & 18.50 & 18.33 \nl
 3 & $-$0.5 & 1.1 & 19.33 & 18.21 & 17.07 & 16.53 \nl
 4 & $-$2.0 & 0.7 & 20.18 & 19.61 & 18.43 & 18.13 \nl
 5 & $-$0.3 & $-$1.1 & 19.97 & 18.74 & 16.88 & 16.55 \nl
\cutinhead{IRAS 20414$-$1651}
 1 & 1.8 & 1.1 & 19.86 & 18.21 & 17.86 & 17.18 \nl
 2 & $-$3.8 & $-$9.0 & 20.69 & 19.76 & $>$ 18.40 & $>$ 18.13 \nl
\cutinhead{IRAS 22491$-$1808}
 1 & 0.0 & 1.0 & 21.44 & 18.46 & 16.98 & 16.65 \nl
 2 & $-$2.5 & 1.1 & 19.79 & 19.38 & 17.18 & 16.73 \nl
 3 & $-$3.2 & 1.5 & 20.35 & 19.38 & 17.19 & 16.73 \nl
 4 & $-$1.0 & $-$0.6 & 19.07 & 18.00 & 16.92 & $>$ 17.19 \nl
 5 & 0.4 & $-$1.1 & 20.28 & 19.11 & 18.56 & $>$ 17.76 \nl
\cutinhead{IRAS 23365+3604}
 1 & $-$1.9 & 1.4 & 19.93 & 18.45 & $>$ 18.00 & $>$ 17.65 \nl
 2 & 2.2 & 1.3 & 21.55 & 20.62 & $>$ 18.88 & $>$ 18.69 \nl
 3 & 1.4 & $-$2.4 & 20.29 & 19.57 & $>$ 18.88 & $>$ 18.25 \nl
 4 & $-$0.3 & $-$1.5 & 19.70 & 17.84 & $>$ 17.80 & $>$ 17.32 \nl
\enddata
\tablenotetext{}{$^1$ RA and DEC offsets measured from the centroid of the ``nucleus'', i.e., the brightest emission source at I-band}
\tablenotetext{}{Reference to the knots should be made in accordance with the I
AU naming convention (galaxy name):SS(number).}
\end{deluxetable}
}


\end{document}